\documentclass[11pt,headings=small,numbers=enddot,twocolumn=on]{scrartcl}
\usepackage{times}
\usepackage{paralist}
\usepackage{csquotes}
\usepackage{overpic}
\usepackage{graphicx}
\usepackage{xcolor}
\usepackage{pdfpages}
\usepackage[numbers,square,sort&compress]{natbib}
\usepackage[american]{babel}
\usepackage{microtype}
\usepackage{tabulary}
\newif\ifblind
\blindtrue
\blindfalse

\newif\ifreview
\reviewtrue
\reviewfalse

\newif\ifcount
\countfalse

\ifcount
\reviewfalse
\KOMAoptions{twocolumn=off}
\else
\ifreview
\countfalse
\KOMAoptions{twocolumn=off}
\usepackage[margin=25mm,right=30mm]{geometry}
\usepackage{setspace}
\usepackage[nolists,noheads]{endfloat}
\renewcommand{\includegraphics}[2][]{}
\usepackage{lineno} 
\linenumbers

\usepackage[none]{hyphenat}
\usepackage{ragged2e}
\RaggedRight\else
\usepackage[left=17mm,right=17mm,top=15mm,bottom=20mm]{geometry}

\usepackage{scrpage2}
\ihead{\normalfont\scshape Bricks and urbanism in the Indus Civilization rise and decline}
\ohead{\normalfont\scshape \ifblind Author \& Author\else Khan \& Lemmen\fi}
\ofoot[\itshape Page~\thepage]{\itshape Page~\thepage}
\ifoot{\normalfont\rmfamily\footnotesize Manuscript to be submitted to \emph{PLOS One}, \today}
\fi\fi

\ifcount\renewcommand\comment[1]{}\fi

\addtokomafont{subsection}{\rmfamily\normalsize\itshape}
\addtokomafont{section}{\rmfamily\normalsize\bfseries}

\newcommand{\reffig}[1]{Fig.~\ref{fig:#1}}
\newcommand{\reffigs}[2]{Figs.~\ref{fig:#1}--\ref{fig:#2}}
\newcommand{\reftab}[1]{Tab.~\ref{tab:#1}}

\title{\large Bricks and urbanism in the Indus Civilization}

\ifblind\else\author{\normalsize Aurangzeb Khan and Carsten Lemmen\thanks{Corresponding author.  Helmholtz-Zentrum Geesthacht,  Max-Planck-Stra{\ss}e~1, 21502~Geesthacht, Germany. Tel.:~+49~4152~87-1521, Fax.:~+49~4152~87-2020, Email.:~carsten.lemmen@hzg.de}\\ \normalsize Helmholtz-Zentrum Geesthacht, Institute of Coastal Research, Germany}\fi

\date{\normalsize\ifcount\else\vspace{\baselineskip}
\begin{minipage}{\hsize}
\noindent\textbf{Abstract.} 
The Indus Civilization, often denoted by its major city Harappa, spanned almost two millennia from 3200 to 1300~BC. Its tradition reaches back to 7000~BC: a 5000~year long expansion of villages and towns, of trading activity, and of  technological advancements culminates between 2600 and 1900~BC in the build-up of large cities, writing, and political authority; it emerges as one of the first great civilizations in history.  During the ensuing 600~years, however,  key technologies fall out of use, urban centers are depopulated, and  people emigrate from former core settlement areas.  Although many different hypotheses have been put forward to explain this deurbanization, a conclusive causal chain has not yet been established.  We here combine literature estimates on brick typology, and on urban area for individual cities.  In the context of the existing extensive data on Harappan artifact find sites and put in their chronological context, the combined narratives told by bricks, cities, and spatial extent can provide a new point of departure for discussing the possible reasons for the mysterious ``decline''.

\end{minipage}
\vspace{-1\baselineskip}
\fi
}

\begin{document}

\ifcount\pagestyle{empty}\else\maketitle\fi
\ifreview\thispagestyle{empty}\fi

\section{Introduction}
%
In the 1850s, ancient bricks stolen from ruins near Harappa, a town adjacent to the River Ravi in Punjab, Pakistan, attracted archaeological investigation.  The bricks were first thought to be part of a Buddhist site, until Marshall \citep{Marshall1924} attributed them to an indigenous civilization of South Asia, the Indus Valley Culture (3200--1300~BC, now more aptly termed the Indus Civilization), whose brick architecture extends back to 7000~BC and the valleys of Baluchistan \citep{Possehl1990,Jarrige1995,Kenoyer1998}.   

The building material for the villages and cities of the Indus Civilization was predominantly mud brick. Only between approximately 2600--1900~BC, in the Mature Harappan phase, were baked bricks used in quantity, especially for walls and floors exposed to water \citep{Possehl2002,Datta2001}.  This period of baked brick usage coincides with an elevated level of urbanism, characterized by large cities as opposed to the predominant village settlements before and after the Mature phase.  In this urban period all other Indus Civilization key technologies, including writing, shell ornaments, weights, and seals are present; these fall out of use with deurbanization after 1900~BC \citep[]{Kenoyer1998,Datta2001,Possehl2002,Law2007}.

At the height of the Indus Civilization, there is thus an intimate relationship between key technologies, building material, and cities.  Or, translated into the social realm, between social and political organization, craftsmanship, and life style.  The investigation of the combined evidence for technologies, material, and cities could then possibly inform us about the social, political, or organizational factors involved in its decline.   We reconstruct here the chronological dynamics of brick usage by typology, and of urban area for the entire Indus domain from individual estimates.   Combined with existing extensive data on Harappan artifact find sites a narrative of the characteristic rural-urban relationship in the Indus Civilization emerges. 

\section{Methods and Material}
\subsection{Chronology and extent}
Several chronologies have been developed for the Indus Cultural Tradition, of which those by Kenoyer \citep{Kenoyer1998} and by Possehl \citep{Possehl2002} are widely employed. From these two, we here differentiate an Early Neolithic period, consisting of the Kili Ghul Muhammad (7000--5000~BC) and Burj Basket-Marked (5000--4300~BC) phases, a pre-Harappan or Developed Neolithic period with the Togau (4300--3800~BC) and Hakra-Kechi (3800--3200~BC) phases, the Indus Civilization proper represented in the Early (3200--2600~BC), Mature (2600--1900~BC), and Late (1900--1300~BC) Harappan phases, and a post-Harappan phase (from 1300~BC).  

As the constituent phases have been developed from local stratigraphies at type sites dispersed throughout the Indus domain, conflicting phase boundaries, multiple naming for cotemporal phases, and phase overlaps, a separation in the eight phases named above is overly simplified and in a way arbitrary; this simplification is necessary, however, to be able to address global developments in a common temporal frame \citep{Fuller2006,Gangal2010\ifblind\else,Lemmen2012har\fi}.  But even if a common temporal frame is established, the calendric ages should be treated with caution, as more often than not precise dating has not been performed for most sites.  Here, we rely on previously published local chronologies (Supplementary Information Table~S1); as the focus of this paper is not on discussing the divergent local chronologies and dating problems, we continue our analysis in the awareness that this chronology is subject to discussion and further refinements in many places.

\begin{figure}[b]
\centering
\includegraphics[width=\ifreview0.8\else\fi\hsize]{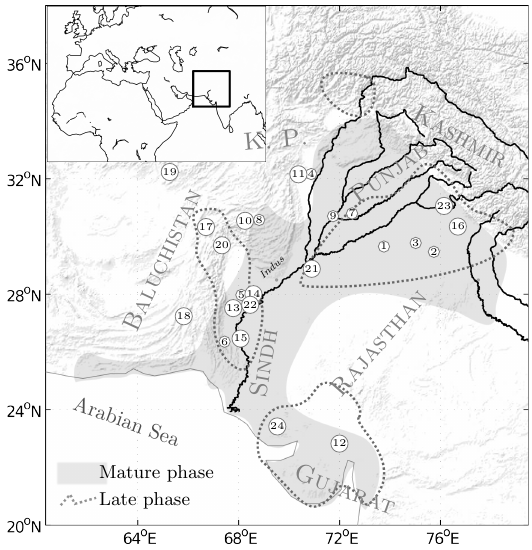}
\caption{Spatial extent of the Indus Civilization in the Mature and Late Harappan phases with  urban centers and villages mentioned in the text: 1~Kalibangan, 2~Rakhigarhi, 3~Banawali, 4~Rehman Dheri, 5~Naru Waro Dharo, 6~Amri, 7~Harappa, 8~Rana Ghundai, 9~Jalilpur, 10~Sur Jangal, 11~Gumla, 12~Lothal, 13~Mohenjodaro, 14~Lakhueenjodaro, 15~Chanhudaro, 16~Bhagwanpura, 17~Damb Sadat, 18~Nindowari, 19~Mundigak, 20~Mehrgarh, 21~Ganweriwala, 22~Kot Diji, 23~Sanghol, and 24~Dholavira.
}
\label{fig:map}
\end{figure}

The Indus Cultural Tradition dates back to around 7000~BC and the foothills and valleys of Baluchistan.  At the site of Mehrgarh (site~20 on map \reffig{map}), early food production was dated to 6500~BC \citep{Jarrige1995}; already early villages exhibit a planned layout, and houses were built of mud bricks.  Pottery appears in the Burj period (after 5000~BC), as well as a wide array of tools, domesticates, and first copper artifacts \citep{Moulherat2002,Fuller2011}.  The occupation area, which had been initially concentrated in Baluchistan, the Makran coast, and the western borderlands of the Indus, expands north- and westward into Khyber Pukhtunkhwa (K.\,P.), Gujarat and the Punjab plains \ifblind(\citep{Gangal2010}; Authors, 2012)\else\citep{Gangal2010,Lemmen2012har}\fi. Use of ornamental pottery and gold emerges,  along with the manufacture of compartmented seals, glazed steatite, and beads. Standardized  weights indicates that trade was important for the pre-Harappan economy.  

The first pre-Harappan cities appear as Mehrgarh, Amri, and Kotdiji  before 3500~BC; they are built from mud (i.e., sun-dried) brick.  Many more villages than cities continue to expand the cultural domain along the Ghaggar Hakra river and along the Makran coast with a doubling of sites numbers after 3200~BC \citep{Law2007}. In this Early Harappan phase baked bricks appear at few sites, first at Kalibangan, Kotdiji, and Banawali \citep{Chattopadhyaya1996}.

At its peak, the mature Indus Civilization extends across the alluvial plains of Punjab and Sindh, Baluchistan, the Gujarat coast, and the surrounding valleys in K.\,P.; in total, it encompasses a vast area of 1~million~km$^{2}$ represented by thousands of individual sites 
(\reffigs{map}{sites}, \citep{Possehl1997,Law2007,Gangal2010}).  Many large cities have been recognized, amongst them are the sites of Harappa and Mohenjodaro with a peak population of approximately 40\,000 inhabitants each \citep{McIntosh2007}; the total population in the Mature phase is estimated at a few million \ifblind(Authors 2012)\else\citep{Lemmen2012har}\fi.  The Harappan extensive and long-range trade network connected by sea to the Sumerian domain and the Arabian peninsula, by land to the Bampur valley and across central Asia \citep{Rao1965,Boivin2009a,Law2011}. 

After 1900~BC, the trade network collapses and weights are disused; large cities are abandoned and baked brick manufacturing discontinues; shell ornament and seal production ceases, and settlements are moved eastwards into the Ganges valley \citep{Possehl2002,Kenoyer1998,Datta2001}.  Sites in Gujarat seem to last longer than all western sites, but by 1300~BC only few scattered sites remain \citep{Rao1963,McIntosh2007}.  The mystery and challenge in Indus scholarship lies in the unresolved causes for and diverse opinions about this decline; most popular theories include environmental change \citep[e.g.,][]{Raikes1964,MeherHomji1973,Staubwasser2003,MacDonald2011},  river relocations \citep[e.g.,][]{Wilhelmy1967,Giosan2012}, or social causes \citep[e.g.,][]{Mackay1948,Wheeler1968,Kenoyer2005}. 
All of these theories, however, suffer from contrasting evidence, interpretation uncertainties, a temporal mismatch with the decline period, or have been reinterpreted to serve a particular political and historical view \citep{Possehl2002,Guha2005}. 
Most certainly, multiple factors contributed to the decline, including  a breakdown of trade and religion \citep{Kenoyer2005}.


\subsection{City and town size through time}
The classification of settlements into villages, towns, and cities is not straightforward for the Indus Civilization, because of the unquantified contribution of pastoralist activities in between settlements \citep{Possehl2002}.  An enclosing wall would classify a settlement as a city, a gated location with political power over the surrounding rural area and a central place for the exchange of traded goods \citep{Kenoyer1998}; a sufficient population size for division of labour and an economic and social organization allowing population growth would identify cities \citep{Childe1950,Modelski2003}.  To a first degree, however, a settlement's absolute area is a proximate measure for its population size.  The relative size of a site compared to the average size of other settlements can then be used to define towns (e.g. 3--7 times larger) or cities (larger than towns).

\begin{table*}[t]
\centering
\caption{Areal extent of all cities and towns where diachronic information was available.}
\label{tab:citysize}
\begin{tabulary}{\hsize}{l  p{118mm} r}\hline
City/Town & \parbox{\hsize}{\raggedleft   Area in hectares (period, year BC)} & Ref.\\[1ex] \hline\\[1ex]
Mehrgarh & \parbox{\hsize}{\raggedleft   30~(7000--5500),    16~(5500--4800),    9~(4800--3500),    9~(3500--3250),    18~(3250--3000)}& \citep{Possehl2002}\\[1ex]
Ganweriwala&\parbox{\hsize}{\raggedleft   79~(2600--2250),    81.5~(2250--1900), 80~(1900--1800)} & \citep{Possehl2002,Law2007,Agrawal1993b}\\[1ex]
Kalibangan &\parbox{\hsize}{\raggedleft   10~(2600--2250), 12~(2250--1900), 5~(1900--1300)} & \citep{Possehl2002}\\[1ex]
Amri&\parbox{\hsize}{\raggedleft  10~(4300--3800), 30~(3800--3300), 20~(3300--3200)}& \citep{Possehl1990}\\[1ex]
Lothal&\parbox{\hsize}{\raggedleft  4.2~(2600--2250), 25~(2250--1900), 25~(1900--1700), 10~(1700--1300)}& \citep{Possehl2002,Rao1963}\\[1ex]
Dholavira& \parbox{\hsize}{\raggedleft 2~(2600--2250), 30~(2550--2200), 80~(2200--2000), 100~(2000--1900), 70~(1900--1800), 15~(1650--1450)}& \citep{Kenoyer1998}\\[1ex]
Harappa&\parbox{\hsize}{\raggedleft  1~(3800--3200), 1.5~(3200--2800), 32~(2800--2600), 100~(2600--2200), 150~(2200--1900), 100~(1900--1800), 8~(1800--1500)}& \citep{Possehl2002,Kenoyer1998}\\[1ex]
Rakhigarhi&\parbox{\hsize}{\raggedleft   2~(3200--2600), 0~(2600--1800), 80~(1800--1300)}& \citep{Possehl2002}\\[1ex]
Mohenjodaro&\parbox{\hsize}{\raggedleft   5~(2900--2600), 75~(2600--2250), 200~(2250--1900), 100~(1900--1700), 20~(1700--1300)}& \citep{Possehl2002,Kenoyer1998,Dani1971}\\[1ex]
\hline
\end{tabulary}

\end{table*}

For nine such cities or towns, chronologically resolved size estimates are available.  These are Mehrgarh, Amri, Kotdiji, Harappa, Mohenjodaro, Lothal, Dholavira, Kalibangan, Ganweriwala, and Rakhigarhi
(\reftab{citysize}).   We generate for each of these cities a continuous time series of area from the earliest to the latest literature estimate.  We calculate the total urban area only from those nine cities whose temporal information is available to show the relative temporal trend in Harappan urban occupation size.  This number underestimates the absolute area, because some other large cities, like Tharowarodar, Nagoor, Nindowari, and Lakhueenjodaro where no temporal trend information is available,  have a combined area of ca. 200\,ha during the Mature Harappan phase \citep{Possehl2002}.  

\subsection{Brick typology}
The first usage of mud bricks worldwide is recorded for Jericho or Tell Aswad,  dated to around 7500~BC \citep{Wright1985,Stordeur2007}.  In the Indus Cultural Tradition, mud bricks at Mehrgarh have been used since around 7000~BC. A recent discussion of possible Near Eastern roots of the Indus Civilization \citep{Gangal2014} provided a cultural spread rate of about 0.65 km per year, much too slow to account for an import of the idea of mud bricks at Mehrgarh, which is more than 4000 km distant from Jericho.

Baked bricks made their first appearance at Jalilpur around 2800~BC \citep{Mughal1970}. Bricks are a signature mark of the entire Indus Cultural Tradition, and baked brick work is the signature mark of its  Bronze Age technologies.   Most of the Indus Civilization's large cities, e.g.\ Harappa, Mohenjodaro, Kot Diji, Ganweriwala, Rakhigarhi, and Lothal have been constructed from both mud and baked brick  (\reftab{brick}, with the largest baked brick to mud brick proportion at Mohenjodaro \citep{Jansen1985}).  Mud brick usage precedes baked brick usage, and continues when baked bricks are not used any more \citep{Datta2001,Chattopadhyaya1996}. Only one large city, Dholavira, is built completely from stones and mud bricks \citep{Bisht1982,Possehl2000}. 

In contrast to the cities, most villages and towns of the Indus Civilization are built from stones and mud bricks \citep{Datta2001,Chattopadhyaya1996}.  The few exceptions are Jalilpur, Kalibangan, and Chanhudaro, where also baked bricks have been used \citep{Mughal1970,Joshi1984,Flam1981}.  Chanhudaro stands out in this list, as there is no preceding mud-brick only phase for this site \citep{Flam1981}.

\begin{table}[t]
\centering
\caption{Brick typology and usage time for all locations where this information was available.  Dates correspond to local stratigraphic chronology (Table S1).}
\label{tab:brick}
\small
\noindent\begin{tabular}{l c c}\hline
Location & Mud brick  & Baked brick \\ \hline
\textit{Urban sites}  &&\\
Harappa \hfill \citep{Datta2001,Chattopadhyaya1996,Possehl2002} &3800--1300&2500--1800\\
Mohenjodaro \hfill \citep{Datta2001,Chattopadhyaya1996} &2900--1300&2600--1800\\
Kot  Diji \hfill \citep{Datta2001,Chattopadhyaya1996,Possehl2002} &3200--1300&2600--1800\\
Banawali \hfill \citep{Datta2001,Bisht1982,Bhan1975}&3200--1000&2600--1800\\
Ganweriwala \hfill \citep{Mughal1997} &2900--1800&2600--1900\\
Rakhigarhi \hfill \citep{Nath2001} &3200--1300&2200--1600\\
Dholavira \hfill \citep{Possehl2002,Bisht1982,Possehl1979} &3200--1450&\\
Lothal \hfill \citep{Datta2001,Rao1979} &2600--1300&2500--1500\\[1ex]
\textit{Non-urban sites} &&\\
Mehrgarh \hfill \citep{Datta2001,Possehl2002,Jarrige1995}&7000--3800&\\
Kili  Ghul  Muhammad \hfill \citep{Datta2001,Fairservis1956} &6000--1800&\\
Sur  Jangal \hfill \citep{Datta2001,Fairservis1959}&4300--3800&\\
Rana  Ghundai \hfill \citep{Datta2001,Chattopadhyaya1996,Lamberg-Karlovsky1972,Stein1929,Ross1946}&4000--1800&\\
Damb  Sadat \hfill \citep{Datta2001,Chattopadhyaya1996,deCardi1983} &3200--2600&\\
Mundigak \hfill \citep{Datta2001,Chattopadhyaya1996,Ball1982} &3300--1800&\\
Amri \hfill \citep{Datta2001,Chattopadhyaya1996,Flam1981}&3200--1800&\\
Chanhudaro \hfill \citep{Possehl2002,Flam1981,Majumdar1934} &&2500--1700\\
Sanghol \hfill \citep{Datta2001,Joshi1984,Bhan1975}&1800--500&\\
Bhagwanpura \hfill \citep{Datta2001,Chattopadhyaya1996,Joshi1984,Bhan1978}&2600--1500&\\
Nal  Samadhi \hfill \citep{Chattopadhyaya1996,Rai1989}&3000--2200&\\
Kalibangan \hfill \citep{Datta2001,Possehl2002,Thapar1979,Joshi1984}&3200--1000&2600--1800\\
Jalilpur \hfill \citep{Mughal1970} &3300--1500&2800--1800\\
Gumla \hfill \citep{Dani1971} &5000--3200&\\
Rehman  Dheri \hfill \citep{Dani1971} &2900--1800&\\ \hline
\end{tabular}
\end{table}

\subsection{Spatiotemporal distribution of sites}
\begin{figure*}[!tb]
\centering
\includegraphics[width=\hsize]{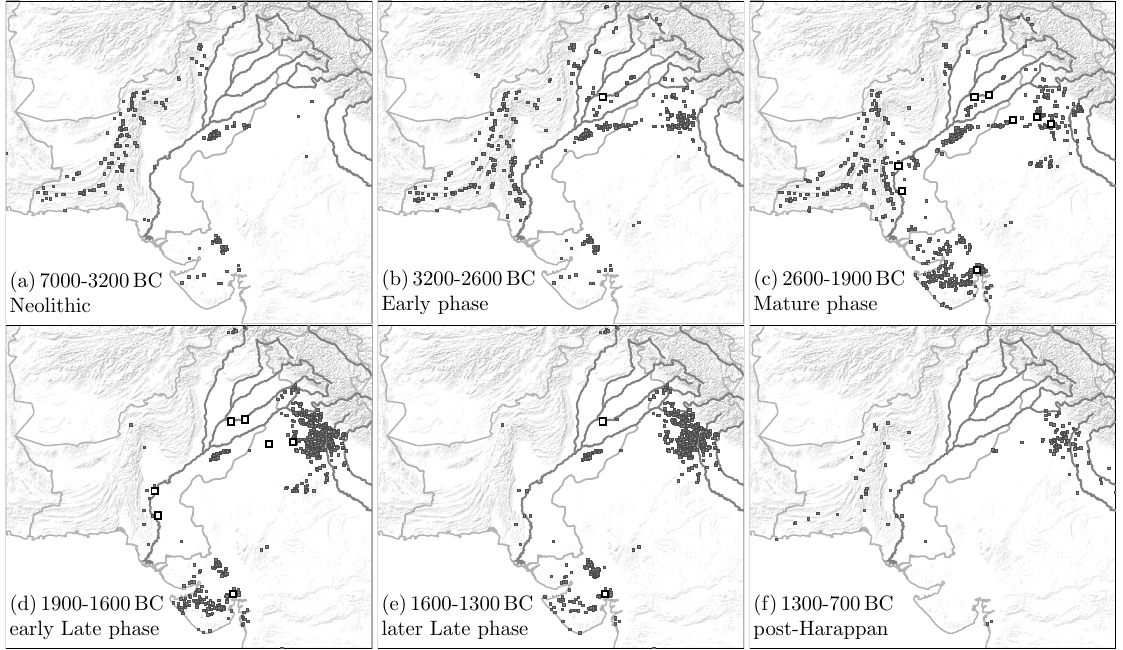}
\caption{Spatial and temporal distribution of find sites of the Indus Cultural Tradition \citep{Law2007} with the chronology from Table S1. Sites with baked brick usage are highlighted with large square symbols.}
\label{fig:sites}
\end{figure*}

We used the Indus Google Earth Gazetteer (version August~2008, \citep{Law2007}) for the geolocation of artifacts relating to the Indus Civilization.  From this database of 3348 dates from 2125 distinct find sites, we use here the cultural attribution and location of those 
3102 Neolithic and Bronze Age sites and dates that are in the spatial and temporal domain of our study.  For the pre-Harappan phases, the Kili Ghul Muhammad, the Burj Basked-Marked, the Togau (and Sheri Khan Tarakai), and the Hakra/Kechi Beg (including Anarta complexes), we extracted 374, 421, 706, and 1039 dates, respectively.  For the Harappan Early, Mature and Late phases, we obtained 1321, 1848, and 1085 dates, respectively (\reffig{sites}).
To our knowledge, this gazetteer is the most extensive and the most representative data set of lithic and metal artifacts of the Indus Cultural Tradition; it overlaps closely with the 3173 sites published by Possehl \citep{Possehl1997}.  This dataset has been used to investigate the trade and distribution networks of the Indus Civilization by its author \citep{Law2011}.   This author cautions that probably 95\% of the sites are not geolocated correctly, but that the error is mostly due to rounding and approximation of geocoordinates by the surveyors, such that the misallocation is only on the order of few or several kilometers;  large-scale analyses such as ours are not affected by these inaccuracies in the data.

\section{Bricks and urbanism in rise and decline}
\subsection{Bricks}
Bricks constitute a large part of the archaeological material left behind by the Indus Civilization.  These bricks provide information on the number and spatial distribution of settlements, on the extent and density of settlements, and on the relation between dense urban and sparse rural areas. Even more, bricks inform about structural and symbolic functions in ancient societies \citep{Kenoyer2006}

Brick work literally lays the foundations of the Indus Cultural Tradition when it emerges after 7000~BC.  Its prominent role, however, is taken by baked bricks, which were manufactured only from the end of the Early to the beginning of the Harappan Late phase, a distinct and narrow 1500~year period within the almost six millennia long tradition.  Why this shift to and away from baked bricks?  

Mud brick was the preferred construction material at the pre-Harappan site of Rana Gundai in Baluchistan despite abundant availability of building stones \citep{Ross1946}.  One functional reason for using mud bricks could have been the better thermal insulation, one aesthetic reason the better sound insolation of mud brick walls  \citep{Binici2007,Doyle2011}.  Mud bricks harden very fast---within one week of exposure to sun, and their utility as a construction material is greatly improved by the addition of straw, which increases the bending and compressive strength and avoids too much shrinkage during the drying process \citep{Binici2007}.

Mud bricks, however, are not as resistant to water and compression as baked bricks  \citep{Lourenco2010,Fernandes2007}.   While most of the building continued to be performed with mud bricks, baked bricks were extensively used where their improved qualities were important \citep{Possehl2002}.  
Water resistance was required for baths, drainage systems and flood protection structures, which are recurrently or permanently exposed to water; water resistance became a key factor in the expansion of Harappan villages and cities into the Punjab flood plains and their sustained establishment in the flooding zones of the river plains was facilitated by baked brick technology.  The protective function of baked bricks is exemplified by the massive and technically refined flood protection structures around Mohenjodaro and Harappa \citep{Kenoyer1998}.  Baked brick usage for all buildings in the flood-prone city Chanhudaro demonstrates the importance of baked-brick technology for flood protection.

The construction of granaries, city walls, and citadels relied on the higher compressive strength of baked brick;  they were used for city walls and citadels in the four largest cities Mohenjodaro, Harappa, Ganweriwala, Rakhigarhi, and several minor cities and purpose sites \citep{Smith2006b}.   The production of baked brick, however, is costly: bricks need to be heated to more than 500 degrees Celsius for several hours to achieve sufficient strength \citep{Papadopoulou2006}. Thus, almost all large cities of the Mature Harappan phase (except Dholavira, Chanhudaro) are built with a combination of sun-dried and baked bricks.  Baked brick structures also needed continuous maintenance: on average, the flood protection lining needed to be replenished at 200~year intervals in the pertaining climate conditions.

The temporal distribution of the number of cotemporal sites using bricks (from \reftab{brick}, shown in \reffig{bricks}a) shows four different dynamic regimes of overall brick usage:  (1)~a steady and slow increase characterizes the Neolithic periods, (2)~a sudden doubling and steep increase of brick sites is typical for the Early Harappan phase; this (3)~levels out during the Mature phase before it is (4)~reversed by a strong decrease during the Late and post-Harappan phases.  Only during the Mature phase, most of the sites investigated use also baked bricks; and baked bricks decline entirely during the Late phase.

The baked brick technology, once invented,  required skilled labor, standards, and natural resources.  All these were available in the Mature Harappan phase.  There is no evidence for scarcity of natural resources for baked brick production.  Fine silt (and water) abounded in the river plains  of Punjab and Sindh.  Irrespective of potential climatic changes, the gallery forests along the perennial rivers provided an ample and steady supply of fire wood: Meher-Homji \citep{MeherHomji1973} estimated that only 200~hectares of riverine forest were required to supply baked bricks long enough to support the large city of Mohenjodaro (which was mostly built from baked bricks) for 100~years. 

The second requirement---standards---has been a long-standing and featured trademark of Harappan masonry. Possehl \citep{Possehl2002} calls the typical ratio of 4:2:1 (length to width to height) of bricks the ``Indus proportion''.  The adherence to this ratio was ensured by the use of standardized molds that have been in use since 4000--3600~BC \citep{Jarrige1995}.  While this ratio was typical at Harappa for large bricks, some cities, like Kalibangan, also used different brick ratios (3:2:1 \citep{McIntosh2007}).  During the Harappan Late phase brick dimensions diverged away from the Indus proportion \citep{Datta2001,McIntosh2007}. 

Beyond the molds,  the standards are also preserved in the craftsmen's tradition and in social norms. The deviation from the standard in the Harappan Late phase could therefore point to a changed social norm, or to the lack of craftsmen to keep up the traditional brick manufacture. This third requirement of skilled labour refers to the craftsmanship and knowledge needed to choose the correct silts, mix the appropriate quantities of silts and water, and find the right temperature and roasting time to produce maximum strength bricks.  Were key skills lost with the migration of craftsmen? There is no direct evidence.  The late appearance of bricks in the Gujarat sites, predominantly Lothal after 2200~BC \citep{Rao1965}, however, could be evidence for increased need of brick producers there, when at  the same time the size of Harappa already started to decrease.  Outside of the Indus domain, baked brick technology appears in Susa (eastern Gulf of Persia), where they are used in monumental construction from 1800~BC \citep{Gallet2006a}.  

\subsection{Urban center}
\begin{figure}[!t]
\centering
\includegraphics[width=\ifreview0.8\fi\hsize]{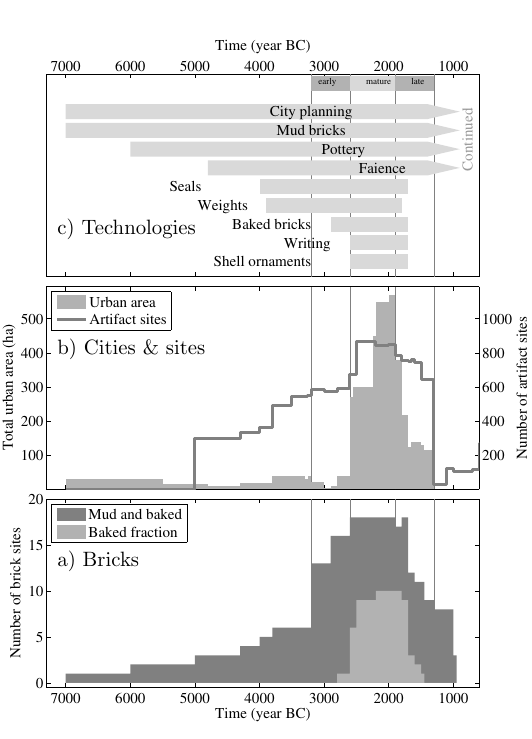}
\caption{(c, bottom) Brick typology in Indus cities, shown as the number of sites employing mud bricks, or employing both mud and baked bricks (\reftab{brick}).
(b, middle) The rural-urban contrast, shown as the total urban area (from \reftab{citysize}) and the number of Indus artifact find sites from the gazetteer \citep{Law2007}.
(c, top) Introduction, presence and expiration of key Harappan technologies \citep{Kenoyer1998,Possehl2002}. Arrows show a continuation of technology in successor cultures.}
\label{fig:bricks}
\label{fig:tech}
\end{figure}

The contrast between rural and urban lifestyle in the Indus Cultural Tradition is best portrayed by the distribution of find sites on the one hand, and by the area of the largest cities on the other hand (\reffig{bricks}b). Between 5000 and 1300~BC, there is a continuous occupation of between 400 and 1000 cotemporal sites recorded in the database \citep{Law2007}.  The increase from pre-Harappan to the Early Harappan phase is around 300~sites, and another increase by 300~sites occurs from the Early to the Mature phase.  The number of sites falls gradually to 800~during the Late phase: in this dataset, there is a slow but not precipitous decline.

It is, however, not the number, but rather the spatial pattern of sites which changes through pre-Harappan \ifblind(Authors 2012)\else\citep{Lemmen2012har}\fi\ and Harappan phases \citep{Joshi1984,Gangal2010}.  At the end of the Mature phase,  only along the Ghaggar-Hakra river sites disappear whereas new sites emerge in the upper Ganges reaches. By 1500~BC,  most of the Baluchistan and Punjab sites have disappeared, while sites in Gujarat and along the Ganges are still present.  The Gujarat complex disappears by 1000~BC  \citep{Gangal2010}; these authors also note that during the terminal Mature phase many small sites replace large sites;  they attribute this to a movement of population from urban centers into villages.

In contrast to the number of (small) sites, the data on large urban centers and their area (\reftab{citysize}) exhibits stronger temporal dynamics.  The combined urban area of large cities was below 40~ha until 2600~BC.  The few pre-Harappan and early Harappan cities (e.g., Kalibangan, Amri Nal) were small, in contrast to the many and large cities of the Mature phase, where the largest cities were Mohenjodaro, Harappa, Ganweriwala, Dholavira, and Rakigarhi, each of them between 80 and 200~ha.  
Total urban area is 450~ha in the first half of the Mature phase, and increases by another 50\% after  2300~BC.  

From the Harappan Early to the Mature, and within the Mature phase, the growth in urban area by far exceeds the slow dynamics of the number of settlements.  This discordance points to an intensification, a population growth within or movement towards the urban centers. Mirroring this intensification, the drastic decrease from 750 in the Mature to 100~ha urban area in the Late phase is not accompanied with a decrease of the site numbers:  the Indus population did not decrease, but rather moved from the cities into smaller (and many) villages

How is the brick dynamics reflected in the urban area?  Both baked bricks and urban centers existed almost exclusively in the Mature Harappan phase, and both experienced a strong increase before and decrease afterwards, pointing to a strong correlation between these two aspects of urbanism.  It is difficult---considering the temporal uncertainties and low number of sites considered---to establish a temporal sequence between the two.  With the current data, the baked brick increase seems to lead the urbanization, which would confirm the role of baked bricks as a prerequisite for urban centers.  The temporal uncertainty in the chronology, however, would need to be decreased for a better quantification of the phase relationship between baked bricks and urban area.

\subsection{Urban mind}
For the Harappan urban mind, baked brick technology represents one of the most important characteristics.  With the baked-brick stimulated rise of urbanism, dense population were possible which stimulated innovation \citep{Lemmen2014boserup};  Mature Harappan phase technologies like writing and shell ornaments flourished (\reffig{bricks}c).   

The Indus Civilization was centrally organized, an empire under common rule. Priestly elites seem to have exerted their power rather by moral authority than force; temples, palaces and evidence for warfare are absent from the archaeological record \citep[][]{Piggott1950,Ratnagar2000}.  The coherence provided by a moral authority may also be a decisive factor in ensuring the brick dimensions standard. Harappa was also a closed society: Skull features from prehistoric cemeteries indicated that urban Harappans differed from surrounding villagers; apparently,  social practice discouraged mixing with people outside the city and promoted endogamy \citep{Bartel1979}\footnote{Even today, endogamy prevails in South Asian populations and preserves genetically distinct tribes \citep{Chaubey2007}}.

\begin{table}
\caption{Simplified mirroring of key Harappan technologies in social relations 
}
\label{tab:mirror}
\centering
\begin{tabular}{l c r}
\hline
Technology && Social relation\\ \hline
Baked bricks & $\leftrightarrow$ & skilled labour\\
Weights & $\leftrightarrow$ & trade links\\
Seals & $\leftrightarrow$ & moral authority\\
Writing & $\leftrightarrow$ & administration\\
Shell ornaments & $\leftrightarrow$ & elite status\\
\hline
\end{tabular}
\end{table}

Skills, trade, authority,  and elite status are social dimensions which can be mapped to the material culture (\reftab{mirror}). The symbolism that held together Harappan society is mirrored in its seals.   Elite status is expressed by shell ornaments.  The Indus script has not been deciphered,  the existence of a writing system, however, points to use for accounting and administration.  Thus, the decline in baked brick manufacturing is not merely a loss of one specific technology, but also represents a considerable loss of symbolism \citep{Possehl2002}.  

\section{Insights for the decline}

Social disruptions at the beginning of the Harappan Late phase put an end to religion and trade.  The changed burial pattern points to a different belief system; additional layers of glaze are used in distinctive pot burials, and glass making and bead drilling techniques were altered.  Trade with the Swat region ceased and resulted in a decline of shell working.    All of this has been interpreted as a disturbance of the elite structure by various authors \citep{Ratnagar2000,Kenoyer2006,Nichols2008}.  Limited evidence points to interpersonal violence among elites, such as postcranial injuries on excavated skulls (Harappa \citep{RobbinsSchug2012}, but not Mohenjodaran \citep{Kennedy1984}), and few cities were (partly) burned, like Kotdiji, Gumla, Nausharo, or Amri \citep{Possehl2002}.  

Environmental changes have been put forward as reasons  for the social upheavals:  Aridity and river relocation long seemed credible for the Ghaggar Hakra area \citep{Staubwasser2003,Giosan2012}, although recent evidence points to possible drying of this river before the Holocene \citep{Sinha2012}. On the other hand, floods frequently occurred in cities in Sindh: a total of 5~m silt, composed of up to 2~m thick individual deposits in Mohenjodaro shows the consequence of inundations from the middle of the Mature to the end of the Late phase; multiple alternating cultural and silt deposit layers indicate rebuilding and renewed floods in this city \citep{Raikes1964,Pruthi2004}.  Even if one ignores the contrasting evidence, the interpretation uncertainties, and the temporal mismatches with the decline period of the suggested environmental factors, there seems to be no environmental reason that could explain the demise of urban centers and relocation of villages from Punjab and Baluchistan. 

Rather, social causes, like an internal political struggle, which lead  to the visible change in the Harappan elite system \citep{Kenoyer2005} should be investigated.  What do bricks tell us about these social changes?  First of all, there is this marked loss of standardization in the Indus brick proportion after 1800~BC, away from the 4:2:1 ratio to 3:2:1 and others, occurring in up to five different transitions until the Iron age \citep{Datta2001,McIntosh2007}.  This divergence could support Kenoyer's theory of a changed elite structure, who had a different architectural preference and who exercised less central control over building practice.  Altered civic control structures could also have resulted in the migration of skilled labour away from the cities (and away from the new administrators), effectively leading to a loss of those skills in the Harappan core settlement area.  Abandoning of urban settlements in favor of smaller villages could have disrupted the balance between the different occupations and the economies of scale needed in the more complicated and costly baked brick making process.  The disintegration of few cities into many small villages would have left many villages entirely without craftsmen trained in baked brick production.

A changed elite may also have not been able, or not have been willing, to continue baked brick manufacturing and the maintenance of the flood protection structures in Punjab and Sindh, rendering those cities vulnerable to flooding at least from 1700~BC. If we take into account evidence that elites were replaced by people originally from outside the Indus domain \citep{Masson1999},  their unfamiliarity with local environmental conditions and protection needs, or their different management priorities could have exacerbated the problem of neglected flood protection.  

Support of a violence theory from bricks can only be speculative at this moment: what if the relocation from the cities to the villages was haphazard, rushed and unplanned, as might be expected from an outbreak of violence in the city?   Refugees would probably not have taken along the bulky molds.  Again, a loss of craftsmanship would be observed, together with a potential loss or diversification of standards like weights and molds.  

Lastly, there is also a circumstantial environmental component to the changes in brick production and morphology:  the general eastward and deurbanization trend was accompanied by the disappearance of straw in the mud bricks; seemingly, the added strengths was either not needed (with smaller construction) or it was much more economical to rebuild with cheap bricks over and over---there are seven construction layers within 200~years in Haryana \citep{Chakrabarti2009}---instead of using the more expensive baked or the more refined straw-filled mud bricks.  The movement of settlements away from the rivers could also have deprived the kilns of firing wood that had been abundant in the riverine gallery forests.  The reason for the deurbanization, however, is to be sought in social reorganization.

\section{Conclusion}
We provide here a novel integrative view of Indus Civilization site distribution, its urban-rural contrast, and the dynamics of brick usage and urban size to find new points of departure for interpreting its decline.  We find that despite a large geographic change of the site distribution, the number of sites and---to first approximation---population does not change much between the Early, Mature, and Late Harappan phases.  Urban area and baked bricks, however, change dramatically in the material culture, as do their social counterparts administration, elite structure, and religion.  By concentrating on the cities, we point to primarily social reasons as a starting point for further investigations on the decline.

\subsubsection*{Acknowledgements}
We are grateful to K.W.~Wirtz for stimulating ideas to and a critical analysis of a prior version of this manuscript.

\bibliographystyle{plos2009}
\bibliography{mendeley-manual}

\ifreview\else

\section*{Supplementary material}

# Bricks and urbanism in the Indus Civilization
# [Supporting Information]

Aurangzeb Khan and Carsten Lemmen*

Helmholtz-Zentrum Geesthacht, Institute of Coastal Research, Germany

## 1. Supporting Tables

Supporting Table S1 summarizes the local chronologies used in this study.

*Corresponding author. Helmholtz-Zentrum Geesthacht, Max-Planck-Straße 1, 21502 Geesthacht, Germany. Tel.: +49 4152 87-1521, Fax.: +49 4152 87-2020, Email.: carsten.lemmen@hzg.de

BRICKS AND URBANISM IN THE INDUS CIVILIZATION RISE AND DECLINE     KHAN & LEMMENTable S1: Summary of local chronologies used in this study

| Chronology | Period (BC) | Chronology | Period (BC) | Chronology | Period (BC) |
|---|---|---|---|---|---|
| Mehrgarh I | 7000–5500 | Early Harappan | 3200–2600 | Bara [9] | 1900–1300 |
| Kili Ghul Moh. | 7000–5000 | Kulli (Early) [5] | 3200–2600 | Jhukar | 1900–1700 |
| Northern Neolithic [1] | 3200-1900 | Kot Diji | 3200–2600 | Late Sorath | 1900–1600 |
| | | Mehrgarh VI | 3000–2800 | Cemetery H | 1900–1500 |
| Mehrgarh II | 5500–4800 | Shahi Tump [4] | 3500–3000 | Rajaput | 1900–1300 |
| Burj Basket-Marked | 5000–4300 | Ahar-Banas [2] | 3000–1500 | Late Harappan | 1900–1300 |
| Mehrgarh III | 4800–3500 | Mehrgarh VIIC | 2800–2600 | Post-urban | 1900–1300 |
| Togau | 4300–3800 | Dasht [6] | 2800–2200 | Prabhas [8] | 1800–1500 |
| Sheri Khan Tarakai | 4300–3000 | Kot Diji (Late) | 2600–1900 | Rangpur IIB [10] | 1800–1500 |
| Anarta [2] | 3500–2600 | Kulli | 2600–1900 | Rangpur IIC [10] | 1800–1500 |
| Hakra Wares | 3800–3200 | Accharwala [7] | 2600–1900 | Pirak II | 1800–1000 |
| Kechi Beg | 3800–3200 | Mature | 2600–1900 | Malwa | 1700–300 |
| Pre-Harappan | 3800–3200 | Harappan | | Swat | 1650–1300 |
| Jodhpura [3] | 2600–1800 | Uttarpradesh [1] | 2600–1900 | Proto-Historic | |
| Mehrgarh V | 3250–3000 | Ganeshwar [3] | 2600–1800 | Lustrous Red Ware | 1600–1300 |
| Nal | 3200–2800 | Quetta [2] | 2500–1900 | Complex B | 1300–700 |
| Anjira [4] | 4300–3200 | Sorath Harappan | 2500–1900 | Jhangar | 1200–1000 |
| Sothi-Siswal | 3200–2600 | Sorath | 2500–1600 | Iron Age | 1200–1000 |
| Amri-Nal | 3200–2600 | BMAC | 2300–1700 | Painted Gray Ware | 1100–500 |
| Ravi | 3200–2600 | Sorathor | 2200–1800 | Pirak III | 1000–700 |
| Damb Sadaat | 3200–2600 | Pre-Prabhas [8] | 2200–1800 | Zangian | 1000–200 |

Based on Possehl [11] and Gangal [12], if not otherwise noted.

## References

[1] Possehl GL (1990) Revolution in the Urban Revolution: The Emergence of Indus Urbanization. Annual Review of Anthropology 19: 261–282.

[2] Fuller DQ (2006) Agricultural Origins and Frontiers in South Asia: A Working Synthesis. Journal of World Prehistory 20: 1–86.

[3] Kenoyer JM (2006) Cultures and Societies of the Indus Tradition. In: Thapar R, editor, Historical Roots in the Making of 'The Aryan', New Delhi: National Book Trust, 1. pp. 10–19.

[4] Franke-Vogt U (2008) Asia, South: Baluchistan and the Borderlands. Encyclopedia of Archaeology .

[5] Biagi P (2011) Changing the prehistory of Sindh and Las Bela coast: twenty-five years of Italian contribution. World Archaeology 43: 523–537.

[6] Mortazavi M (2011) The Bampur Valley: A New Chronological Development. Ancient Asia 1: 53–61.

[7] Possehl GL (1999) Indus Age: The Beginnings. Univ. Of Pennsylvania Press.

[8] Dhavalikar M (1984) Toward an Ecological Model for Chalcolithic Central and Western India. Journal of Anthropological Archaeology 3: 133–158.

[9] Shaffer J (1981) The Protohistoric Period in the Eastern Panjab: A Preliminary Assessment. In: Dani A, editor, Indus Civilization New Perspective, Karachi, Pakistan: East and West. p. 137.

[10] Meadow RH, Allchin FR, Erdosy G, Coningham RaE, Chakrabarti DK, et al. (1997) The Archaeology of Early Historic South Asia: The Emergence of Cities and States. Journal of Asian Studies 56: 514.

[11] Possehl GL (2002) The Indus civilization: a contemporary perspective. Walnut Creek, CA: AltaMira, 276 pp.

[12] Gangal K, Vahia MN, Adhikari R (2010) Spatio-temporal analysis of the Indus urbanization. Current Science 98: 846–852.Manuscript to be submitted to *PLOS One*, July 24, 2014     *Page 2*

\fi

\end{document}